\documentclass[final,5p,fleqn]{elsarticle}

\usepackage{graphicx}
\usepackage{hyperref}
\usepackage{amsmath}
\usepackage{url}
\usepackage{colordvi}
\usepackage{color}

\usepackage[normalem]{ulem} 
\renewcommand\sout{\bgroup \color[rgb]{0.55,0.00,0.99} \ULdepth=-.5ex \ULset}

\begin{document}

\begin{frontmatter}

\title{On the Rotational Invariance and Non-Invariance of Lepton Angular 
Distributions in Drell-Yan and Quarkonium Production}

\author[a]{Jen-Chieh Peng}
\author[b]{Dani\"el Boer}
\author[c]{Wen-Chen Chang}
\author[a,d]{Randall Evan McClellan}
\author[e]{Oleg Teryaev}

\address[a]{Department of Physics, University of Illinois at
Urbana-Champaign, Urbana, Illinois 61801, USA}
\address[b]{Van Swinderen Institute for Particle Physics and
Gravity, University of Groningen, Groningen, The Netherlands}
\address[c]{Institute of Physics, Academia Sinica, Taipei 11529, Taiwan}
\address[d]{Thomas Jefferson National Accelerator Facility,
Newport News, VA 23606, USA}
\address[e]{Bogoliubov Laboratory of Theoretical Physics,
JINR, 141980 Dubna, Russia}


\begin{abstract}

Several rotational invariant quantities for the lepton angular
distributions in Drell-Yan and quarkonium production were
derived several years ago, allowing the comparison between 
different experiments adopting different reference frames.
Using an intuitive picture for describing the lepton angular
distribution in these processes, we show how the rotational
invariance of these quantities can be obtained. This
approach can also be used to determine the rotational invariance
or non-invariance of various quantities
specifying the amount of violation for the Lam-Tung
relation. While the violation of the Lam-Tung relation is often
expressed by frame-dependent quantities, we note that
alternative frame-independent quantities are preferred.
 
\end{abstract}

\end{frontmatter}

The angular distributions of leptons produced in the Drell-Yan
process~\cite{drell} and the quarkonium production in hadron-hadron
collisions~\cite{biino,braaten} remain a subject of considerable interest.
The polar and azimuthal angular distributions of leptons produced
in unpolarized and polarized Drell-Yan process allow the extraction
of various types of transverse-momentum dependent 
distributions~\cite{barone,peng14}.
First (leading order) results on the extraction of the Boer-Mulders 
functions~\cite{boer98,boer99} have been
obtained from unpolarized Drell-Yan experiments using 
pion~\cite{falciano86,conway} or proton~\cite{zhu}
beams, indicating that the quark transverse spin is correlated with 
the quark transverse momentum 
inside unpolarized protons. 
A more precise determination of the amount of quark polarization requires 
inclusion of higher order perturbative corrections because gluon 
radiation can also affect the lepton angular 
distributions \cite{Brandenburg:1993cj,peng16,Lambertsen:2016wgj,chang17}.
Recent measurement of Drell-Yan angular distributions with a pion beam 
on a transversely polarized proton
target provided the first information from Drell-Yan on the 
correlation between the quark transverse momentum and the spin
direction of a transversely
polarized proton~\cite{compass}. For quarkonium production,
the lepton angular distributions reveal sensitively the 
underlying partonic
mechanisms, as various subprocesses could lead to distinct polarizations
for the quarkonium~\cite{braaten,kniehl,chao}. 

The lepton angular distributions in Drell-Yan and quarkonium production
are generally measured in the rest frame of the dileptons. Many different
choices of the reference frames exist in the literature, depending on
how the axes of the coordinate system are chosen. While it is common
to define the $y$ axis to be along the direction normal to the reaction plane
(which is the plane containing the beam axis and the dilepton's momentum
vector) and the $x$ and $z$ axes lying on the reaction plane, the specific
direction of the $z$ axis is chosen differently for different reference
frames. In particular, the Collins-Soper frame~\cite{cs} 
has the $z$ axis bisecting
the beam and target momentum vectors, while the helicity frame aligns the $z$
axis with the dilepton momentum vector in the center-of-mass frame. The
Gottfried-Jackson frame~\cite{gottfried} and 
the $u$-channel frame have the $z$ axis parallel to the
beam and target momentum direction, respectively. These various reference
frames are related to each other by rotations along the $y$ axis by
certain angles~\cite{falciano86,faccioliepj}.

A general expression for the lepton angular distribution in the Drell-Yan 
process or quarkonium production is given as
\begin{eqnarray}
\frac{d\sigma}{d\Omega} \propto 1+ \lambda_\theta \cos^2\theta
+ \lambda_{\theta \phi} \sin 2 \theta\cos\phi
+ \lambda_\phi \sin^2\theta \cos 2 \phi,
\label{eq:eq1}
\end{eqnarray}
where $\theta$ and $\phi$ refer to the polar and azimuthal angles of $l^-$
($e^-$ or $\mu^-$) in the rest frame of the dilepton. While the polar
angle dependence is specified by the parameter $\lambda_\theta$, the
azimuthal dependencies of the lepton angular distributions are described 
by the parameters $\lambda_{\theta \phi}$ and 
$\lambda_\phi$. 
Note that these parameters are related to the
parameters $\lambda, \mu, \nu$ in Ref.~\cite{lam78} as
$\lambda_\theta = \lambda, \lambda_{\theta \phi} = \mu$ and $\lambda_\phi
= \nu/2$.
The values of $\lambda_\theta,\lambda_{\theta \phi}$ and 
$\lambda_\phi$ depend on the choice of 
the coordinate system. While the Collins-Soper frame is chosen
by many experiments for the data analysis, other 
reference frames are also
utilized by some experiments. Going from one frame to another acts as 
a nonlinear transformation on these three parameters~\cite{Boer:2006eq}, 
making it hard to connect the results in different frames.

The frame-dependence of the angular
distribution parameters could potentially lead to confusion when
comparing results of lepton angular distributions or quarkonium 
polarizations measured in different experiments~\cite{faccioliepj,faccioli1}. 
In order to mitigate the confusion caused by the frame dependence of the
parameters $\lambda_\theta$, $\lambda_{\theta \phi}$ 
and $\lambda_\phi$, 
Faccioli et al.~\cite{faccioli2,faccioli3,faccioli4} pointed out that
various quantities can be formed from 
$\lambda_\theta$, $\lambda_{\theta \phi}$
and $\lambda_\phi$
with the
property that they are invariant under the transformations among different
reference frames. The comparison between measurements obtained
with different
reference frames could be performed, if such rotation invariant
quantities are used rather than 
the individual $\lambda_\theta$, $\lambda_{\theta \phi}$
and $\lambda_\phi$ parameters.
Examples of such rotational invariant quantities 
include~\cite{faccioli3,faccioli4} 
\begin{eqnarray}
\mathcal{F}=\frac{1+\lambda_\theta+2\lambda_\phi}{3+\lambda_\theta},
\label{eq:eq2}
\end{eqnarray}
and
\begin{eqnarray}
\tilde \lambda = \frac{\lambda_\theta +3\lambda_\phi} {1-\lambda_\phi}.
\label{eq:eq3}
\end{eqnarray}
The reason for considering these particular combinations is not just 
the rotational invariance, but also that they are measures for the deviation 
of the Lam-Tung relation~\cite{lam80}, 
$1-\lambda_\theta = 4 \lambda_\phi$, that is 
satisfied in the Drell-Yan 
process at order $\alpha_s$ in case of collinear parton distributions. 
Its violation 
results from the acoplanarity of the partonic subprocess, 
as discussed in detail in Refs.~\cite{peng16,chang17}.
This acoplanarity can arise from intrinsic transverse momentum of quarks inside
the proton, but also from perturbative gluon radiation beyond order $\alpha_s$. 
They lead to a deviation 
of $ \mathcal{F}$ from $\frac{1}{2}$ and of $\tilde \lambda$ 
from 1. In contrast, 
the deviation of $1-\lambda_\theta-4\lambda_\phi$ from 
zero often considered in 
experimental and 
theoretical studies~\cite{boer99,falciano86,conway,zhu,Brandenburg:1993cj} 
is {\it not} a rotationally invariant 
quantity, pointed out first in Ref.~\cite{faccioli4}, and 
hence a potential source of confusion when comparing its values 
obtained in different frames. 

Another rotation-invariant quantity invoking all three parameters
is~\cite{braaten,palestini11}
\begin{eqnarray}
\tilde \lambda^\prime = \frac{(\lambda_\theta - \lambda_\phi)^2 
+ 4 \lambda_{\theta \phi}^2} 
{(3 + \lambda_\theta)^2}.
\label{eq:eq3p}
\end{eqnarray}
Although not immediately obvious from their definition in terms of 
$\lambda_\theta$, $\lambda_{\theta \phi}$
and $\lambda_\phi$,
the above three quantities, $\mathcal{F}, \tilde \lambda$, 
$\tilde \lambda^\prime$, are invariant only
under rotations around the $y$ axis, which includes the transformations
connecting the various references frames in the literature. On 
the other hand,
the quantity $\mathcal{G}$ is invariant under
the rotation along the $x$ axis~\cite{faccioli4},
\begin{eqnarray}
\mathcal{G}=\frac{1+\lambda_\theta-2\lambda_\phi}{3+\lambda_\theta}.
\label{eq:eq4}
\end{eqnarray}
Finally, $\lambda_\theta$ is invariant under the rotation along 
the $z$-axis~\cite{faccioli4}.

The rotational invariance of $\mathcal{F}$, $\tilde \lambda$, 
$\tilde \lambda^\prime$ and  
$\mathcal{G}$ was obtained in Refs.~\cite{faccioli3,faccioli4,palestini11,qiu} 
from the consideration of the
covariance properties of angular momentum eigenstates of a vector
meson. In a recent study~\cite{peng16,chang17}, it was shown that 
some salient features of the
parameters  $\lambda_\theta$, $\lambda_{\theta \phi}$
and $\lambda_\phi$ in the Drell-Yan process
and $Z$-boson production can be well described by
an intuitive approach. In particular, the 
pronounced transverse-momentum dependence of $\lambda_\theta$ and 
$\lambda_\phi$
for $Z$-boson production and the clear violation of the Lam-Tung
relation at the 
LHC~\cite{cms,atlas} can be well described by this approach.
In this paper, we show how the rotational invariance properties of 
$\mathcal{F}$, $\tilde \lambda$, $\tilde \lambda^\prime$ and 
$\mathcal{G}$ can be deduced using
the approach of Refs.~\cite{peng16,chang17}. It is also clear 
from the analysis below that the rotational invariance
or non-invariance of various quantities
characterizing the violation of the Lam-Tung
relation can be obtained.

\begin{figure}[tb]
\includegraphics*[width=\linewidth]{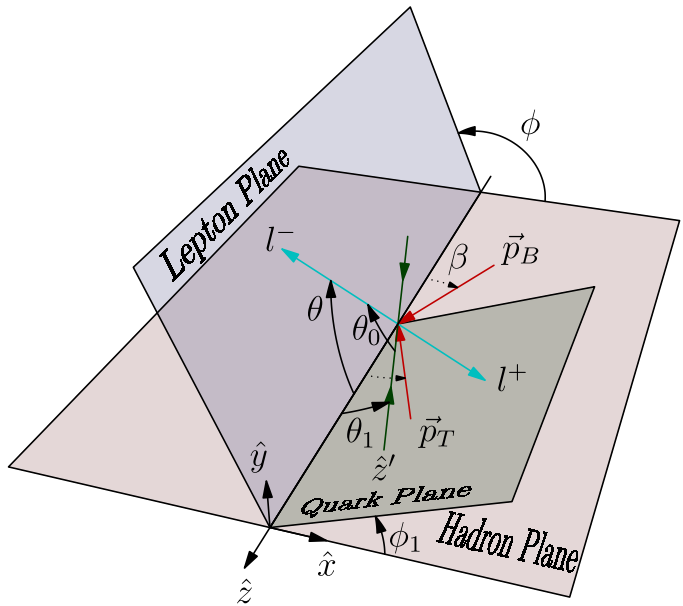}
\caption{Definition of the Collins-Soper frame and various angles
and planes in the rest frame of $\gamma^*/Z$ or a vector quarkonium. 
The hadron plane
is formed by $\vec P_B$ and $\vec P_T$, the momentum vectors of
the two interacting hadrons. The $\hat x$ and $\hat z$ axes
of the Collins-Soper frame both lie in the hadron plane with
$\hat z$ axis bisecting the $\vec P_B$ and $- \vec P_T$ vectors.
The quark ($q$) and antiquark ($\bar q$) annihilate collinearly
with equal momenta to form $\gamma^*/Z$ or a vector quarkonium, 
while the quark momentum
vector $\hat z^\prime$ and the $\hat z$ axis form the quark plane.
The polar and azimuthal angles of $\hat z^\prime$ in the Collins-Soper
frame are $\theta_1$ and $\phi_1$. The $l^-$ and $l^+$ are emitted
back-to-back with $\theta$ and $\phi$ as the polar and azimuthal angles
for $l^-$.}
\label{fig1}
\end{figure}

In the dilepton rest frame, we first define three different planes, namely, the
hadron plane, the quark plane, and the lepton plane,
shown in Fig.~\ref{fig1}. For dileptons with  non-zero transverse 
momentum, $q_T$, the momenta of the two interacting hadrons,
$\vec P_B$ and $\vec P_T$, are
not collinear in the rest frame of $\gamma^*/Z$, and they
form the ``hadron plane" shown in Fig.~\ref{fig1}.
Figure~\ref{fig1} also shows the ``lepton plane", formed by the
momentum vector of the $l^-$ and the $\hat z$ axis. 
In the rest frame of the
dilepton, the $l^-$ and $l^+$ are emitted back-to-back
with equal momenta.

In the dilepton rest frame, a pair of collinear $q$ and $\bar q$
with equal momenta annihilate into a $\gamma^*/Z$ or a vector quarkonium, 
as illustrated in Fig.~\ref{fig1}.
We define the momentum unit vector of $q$ as $\hat z^\prime$, and the
``quark plane" is formed by the $\hat z^\prime$ and $\hat z$ axes.
The polar and azimuthal angles of the $\hat z^\prime$ axis in
the Collins-Soper frame are denoted as $\theta_1$
and $\phi_1$. For the coplanar case, $\phi_1=0$ and the hadron plane
coincides with the quark plane. When $\phi_1 \ne 0$, $\phi_1$ signifies the
acoplanarity angle. With respect to the $q -\bar q$ axis, called the 
natural axis~\cite{Teryaev:2011zza}, the $l^-$ has an azimuthally 
symmetric angular distribution, namely,
\begin{equation}
\frac{d\sigma}{d\Omega} \propto  1 + a \cos \theta_0 + \lambda_0 
\cos^2\theta_0,
\label{eq:eq5}
\end{equation}
where $\theta_0$ is the polar angle between the $l^-$
momentum vector and the $\hat z^\prime$ axis (see Fig.~\ref{fig1}),
and $a$ is the
forward-backward asymmetry originating from the parity-violating
coupling, which is important only when the dilepton mass is close
to the $Z$ boson mass. The parameter $\lambda_0$ depends on the
reaction mechanism. For Drell-Yan process in which a virtual photon
decays into a lepton pair, we have $\lambda_0=1$. This is a consequence
of helicity conservation leading to a transversely polarized virtual photon
with respect to the natural axis. For quarkonium production, the value
of $\lambda_0$ depends on the specific mechanism.
We note that $\lambda_0 = 0$ for unpolarized quarkonium production, 
while $\lambda_0 = -1$ for production of longitudinally polarized 
quarkonium.
   
The angles $\theta$ and $\phi$ are experimental observables, and it
is necessary to express $\theta_0$ in terms of $\theta$ and $\phi$.
This can be accomplished using the following trignometric relation:
\begin{equation}
\cos \theta_0 = \cos \theta \cos \theta_1 + \sin \theta \sin \theta_1
\cos (\phi - \phi_1).
\label{eq:eq6}
\end{equation}
Substituting Eq.~(\ref{eq:eq6}) into Eq.~(\ref{eq:eq5}), we obtain
\begin{eqnarray}
\frac{d\sigma}{d\Omega} & \propto & (1+\frac{1}{2}\lambda_0\sin^2\theta_1) +
(\lambda_0 -\frac{3}{2}\lambda_0 \sin^2\theta_1) \cos^2\theta \nonumber \\
& + & (\frac{1}{2} \lambda_0 \sin 2\theta_1 \cos \phi_1)
\sin 2\theta \cos\phi \nonumber \\
& + & (\frac{1}{2} \lambda_0 \sin^2\theta_1 \cos 2\phi_1)
\sin^2\theta \cos 2\phi \nonumber \\
& + & (a \sin \theta_1 \cos \phi_1) \sin\theta \cos\phi
+ (a \cos \theta_1) \cos\theta \nonumber \\
& + & (\frac{1}{2} \lambda_0 \sin^2\theta_1 \sin 2\phi_1) 
\sin^2\theta \sin 2\phi
\nonumber \\
& + & (\frac{1}{2} \lambda_0 \sin 2\theta_1 \sin\phi_1) \sin 2\theta \sin\phi
\nonumber \\
& + & (a \sin\theta_1 \sin\phi_1) \sin\theta \sin\phi.
\label{eq:eq7}
\end{eqnarray}

A comparison between Eq.~(\ref{eq:eq1}) and Eq.~(\ref{eq:eq7}) shows
that $\lambda_\theta$, $\lambda_{\theta \phi}$, and 
$\lambda_\phi$ can be expressed as a function of 
$\lambda_0$, $\theta_1$ and $\phi_1$ (c.f. with~\cite{Teryaev:2011zza} 
for zero acoplanarity angle $\phi_1=0$):
\begin{align}
\lambda_\theta &= \frac{2\lambda_0 - 3 \lambda_0 \sin^2\theta_1}
{2+ \lambda_0 \sin^2\theta_1} \nonumber \\
\lambda_{\theta \phi} &= \frac{\lambda_0 \sin 2\theta_1\cos \phi_1}
{2+ \lambda_0 \sin^2\theta_1} \nonumber \\
\lambda_\phi &= \frac{\lambda_0 \sin^2\theta_1\cos 2 \phi_1}
{2+ \lambda_0 \sin^2\theta_1}.
\label{eq:eq8}
\end{align}

The terms proportional to $\sin 2\phi$ and $\sin \phi$
do not appear due to Lorentz invariance, provided there are no 
vectors (like transverse polarization) normal to the hadron plane.
Such terms in Eq.~(\ref{eq:eq7}) integrate to zero due to the acoplanarity 
angle average. Unless one considers polarized leptons, parity
or time-reversal violation, Eq.~(\ref{eq:eq7}) reduces to Eq.~(\ref{eq:eq1}).

First, we consider the quantity $\mathcal{F}$ in Eq.~(\ref{eq:eq2}). From
Eq.~(\ref{eq:eq8}), we obtain
\begin{eqnarray}
\mathcal{F}=\frac{1+\lambda_0-2\lambda_0\sin^2\theta_1 \sin^2 \phi_1}
{3+\lambda_0} = \frac{1+\lambda_0-2\lambda_0 y^2_1}{3+\lambda_0},
\label{eq:eq9}
\end{eqnarray}
where $y_1 = \sin \theta_1 \sin \phi_1$ is the component of the unit
vector $\hat z^\prime$ along the $y$-axis in the dilepton rest frame.
The invariance of $\mathcal{F}$ with respect to a rotation along the 
$y$ axis is 
clearly shown in Eq.~(\ref{eq:eq9}), since $\lambda_0$ and $y_1$ are
both invariant under such a rotation. It is interesting to note that
for the Drell-Yan process, where $\lambda_0 = 1$, $\mathcal{F}$ becomes
$(1-y_1^2)/2$. As pointed out in Refs.~\cite{peng16,chang17}, $y_1$, or
the non-coplanarity angle $\phi_1$ between the hadron and the quark planes in
Fig.~1, is in general not equal to zero. For the special case of
$\phi_1 =0$ (or $y_1 = 0$), $\mathcal{F} = 1/2$ and $\mathcal{F}$ is
invariant under any arbitrary rotation in the dilepton's rest frame.   
As discussed in Refs.~\cite{peng16,chang17}, the Lam-Tung relation 
in the Drell-Yan process 
is satisfied when the angle $\phi_1$ vanishes. This is verified 
from Eq.~(\ref{eq:eq8}),
when the values of $\lambda_0$ and $\phi_1$ are set at 1 and 0, respectively.

We next consider the quantity $\tilde \lambda$. Using Eq.~(\ref{eq:eq8}),
Eq.~(\ref{eq:eq3}) becomes
\begin{eqnarray}
\tilde \lambda =\frac{\lambda_0+3\lambda_0\sin^2\theta_1 \sin^2 \phi_1}
{1+\lambda_0\sin^2\theta_1 \sin^2\phi_1} = \frac{\lambda_0+3\lambda_0 y^2_1}
{1+\lambda_0 y^2_1}.
\label{eq:eq10}
\end{eqnarray}
Again, $\tilde \lambda$ must be invariant under a rotation
along the $y$ axis, since $\lambda_0$ and $y_1$ are both invariant
under such rotation. In the
special case of coplanarity between the hadron plane and the quark plane,
we have $y_1 = 0$, and Eq.~(\ref{eq:eq10}) becomes 
$\tilde \lambda = \lambda_0$. In that case, $\tilde \lambda$ 
is invariant
under rotation along any axis.  However, $\tilde \lambda$ is in general not 
the same as $\lambda_0$, and $\tilde \lambda$ is in general not invariant 
under an arbitrary rotation.

We turn our attention next to the quantity $\tilde \lambda^\prime$ in
Eq.~(\ref{eq:eq3p}). All three parameters, 
$\lambda_\theta$, $\lambda_{\theta \phi}$, and
$\lambda_\phi$
are involved in $\tilde \lambda^\prime$. Using Eq.~(\ref{eq:eq8}), we 
obtain
\begin{eqnarray}
\tilde \lambda^\prime =\frac{\lambda_0^2(z_1^2+x_1^2)^2}
{(3+\lambda_0)^2} =\frac{\lambda_0^2(1-y_1^2)^2}
{(3+\lambda_0)^2},
\label{eq:eq10p}
\end{eqnarray}
where $z_1$ is the component of the unit vector $\hat z^\prime$ along the
$z$ axis and the identity $x_1^2+y_1^2+z_1^2 = 1$ is used. 
Thus, $\tilde \lambda^\prime$ is invariant under a rotation
along the $y$ axis. For the coplanar case, $y_1=0$ and $\tilde \lambda^\prime$
is invariant under rotation along any axis.

In an analogous fashion, one can show the invariance of $\mathcal{G}$
and $\lambda_\theta$ under the rotation along 
the $x$ and $z$ axis, respectively.
Using Eq.~(\ref{eq:eq8}), Eq.~(\ref{eq:eq4}) becomes
\begin{eqnarray}
\mathcal{G} = \frac{1+\lambda_0-2\lambda_0\sin^2\theta_1 \cos^2 \phi_1}
{3+\lambda_0} = \frac{1+\lambda_0-2\lambda_0 x^2_1}
{3+\lambda_0},
\label{eq:eq11}
\end{eqnarray}
where $x_1 = \sin \theta_1 \cos \phi_1$ is the component of the unit
vector $\hat z^\prime$ along the $x$ axis in the dilepton rest frame.
Similarly, from Eq.~(\ref{eq:eq8}), the parameter 
$\lambda_\theta$ can be written as
\begin{eqnarray}
\lambda_\theta = \frac{-\lambda_0+3\lambda_0\cos^2\theta_1}
{2+\lambda_0-\lambda_0\cos^2\theta_1} = 
\frac{-\lambda_0+3\lambda_0z^2_1}
{2+\lambda_0-\lambda_0z^2_1},
\label{eq:eq12}
\end{eqnarray}
where $z_1 = \cos\theta_1$ is the component of the unit
vector $\hat z^\prime$ along the $z$ axis in the dilepton rest frame.
From Eq.~(\ref{eq:eq11}) and Eq.~(\ref{eq:eq12}) we note that $\mathcal{G}$
and $\lambda_\theta$ are invariant under 
the rotation along the $x$ and $z$
axis, respectively.

Using the above results one can see that despite the nonlinear transformation of
$\lambda_\theta, \lambda_{\theta \phi}$ and $\lambda_\phi$ 
under rotations, the linear combination
$1-\lambda_\theta - 4\lambda_\phi$ remains zero in 
all other rotated frames if it is zero in one particular frame,
as was observed for specific rotations in~\cite{Boer:2006eq}.
If the combination is nonzero however, 
then its value will change under rotations,
even around the $y$ axis. From Eq.~(\ref{eq:eq8}), it 
follows that the quantity
$1 - \lambda_\theta - 4 \lambda_{\theta \phi}$ is not 
invariant under rotations along the $y$ axis.
On the other hand, the quantity, 
$(1-\lambda_\theta - 4 \lambda_\phi)/(3 + \lambda_\theta)$,
{\it is} invariant under such rotations, namely
\begin{eqnarray}
\frac{1-\lambda_\theta - 4 \lambda_\phi}
{3 + \lambda_\theta} = 1-2\mathcal{F}= 
\frac{1 - \lambda_0 + 4 \lambda_0 y_1^2}{3 + \lambda_0}.
\label{eq:eq13}
\end{eqnarray}
Therefore, to
examine the amount of the violation of the Lam-Tung relation, the 
quantity, 
$(1-\lambda_\theta - 4 \lambda_\phi)/(3 + \lambda_\theta)$,
is preferred. 

Often in the literature for the Drell-Yan process, another set of 
angular coefficients are considered: $A_0, A_1,A_2$, where
\begin{eqnarray}
\frac{d\sigma}{d\Omega} & \propto & (1+\cos^2\theta)+\frac{A_0}{2}
(1-3\cos^2\theta)+A_1 \sin 2 \theta\cos\phi \nonumber \\
& + & \frac{A_2}{2} \sin^2\theta \cos 2 \phi.
\label{eq:eq14}
\end{eqnarray}
The Lam-Tung relation is then expressed as $A_0=A_2$. The violation 
of the Lam-Tung relation, 
$A_0-A_2 = 2(1-2\mathcal{F})$, is rotationally invariant 
around the $y$ axis. On the other hand, 
the quantity $\Delta_{LT} = 1-A_2/A_0$ of \cite{Gauld:2017tww} is not.  

In conclusion, we have presented an intuitive derivation for
rotation-invariant quantities for lepton angular distributions
in Drell-Yan and vector quankonium production. By expressing these
quantities in terms of the $\lambda_0$ and the $x$, $y$ and $z$ components of 
the unit vector of the quark momentum in the dilepton rest frame, the
invariant properties of these quantities become transparent. This
approach offers a useful insight regarding the roles of $\lambda_0$ and
the acoplanarity of the partonic subprocesses in determining 
the applicability and
values of these invariant quantities. 
This approach could also be
extended to other hard processes, such as hadron pair production in 
$e^+ e^-$ annihilation, which is closely connected to the Drell-Yan and
vector quarkonium production.

This work
was supported in part by the U.S. National Science Foundation and
the Ministry of Science and Technology of Taiwan. It was also
supported in part by the U.S. Department of Energy, Office of Science, 
Office of Nuclear Physics
under contract DE-AC05-060R23177.

\end{document}